\documentstyle{mn} 
\input{epsf}
\newif\ifAMStwofonts

\title{Improved synthetic spectra of helium-core white dwarf stars}

\author[Rohrmann et al.]
{R. D. Rohrmann$^{1,2}$\thanks{Postdoctoral Fellow of the IA-UNAM},
A. M. Serenelli$^{3}$\thanks{Fellow of the Consejo Nacional de
Investigaciones Cient\'{\i}ficas y T\'ecnicas (CONICET), Argentina.},
L. G. Althaus$^{3}$\thanks{Member of the Carrera del Investigador
Cient\'{\i}fico y Tecnol\'ogico, CONICET, Argentina.} and
O. G. Benvenuto$^{3}$\thanks{Member  of  the  Carrera  del Investigador
Cient\'{\i}fico, Comisi\'on  de Investigaciones  Cient\'{\i}ficas
de la Provincia de Buenos Aires, Argentina.} \\
$^1$Observatorio Astron\'omico, Universidad Nacional de
C\'ordoba, Laprida 854, (5000) C\'ordoba, Argentina\\
$^2$Present address: Instituto de Astronom\'{\i}a, UNAM, 
A.P. 70-264, 04510 M\'exico D.F., M\'exico\\
$^3$Facultad de Ciencias Astron\'omicas y Geof\'{\i}sicas, Universidad
Nacional de La Plata, Paseo del Bosque S/N, (1900) La Plata, Argentina \\
E-mail: rohr@astroscu.unam.mx,
serenell,althaus,obenvenu@fcaglp.fcaglp.unlp.edu.ar}

\date{2002 April 15}

\pagerange{\pageref{firstpage}--\pageref{lastpage}}

\pubyear{2002}

\begin{document}

\maketitle

\label{firstpage}

\begin{abstract} We examine the emergent fluxes from helium-core white
dwarfs  following their  evolution  from the  end  of pre-white  dwarf
stages down to advanced cooling  stages.  For this purpose, we include
a detailed treatment of the physics of the atmosphere, particularly an
improved representation of the state of the gas by taking into account
non-ideal  effects according to  the so-called  occupation probability
formalism.   The present calculations  also incorporate  hydrogen line
opacity  from  Lyman,  Balmer  and  Paschen  series,  pseudo-continuum
absorptions   and   new   updated   induced-dipole   absorption   from
H$_2$-H$_2$, H$_2$-He and H-He pairs.

We  find that  the non-ideal  effects  and line  absorption alter  the
appearance of  the stellar spectrum  and have a  significant influence
upon the photometric  colours in  the UBVRI-JHKL  system.  This occurs
specially  for  hot models  $T_{\rm  eff}\ga  8000$  due to  line  and
pseudo-continuum opacities, and for  cool models $T_{\rm eff}\la 4000$
where the  perturbation of atoms  and molecules by  neighbour particles
affects the chemical equilibrium of the gas.

In  the present  study,  we  also include  new  cooling sequences  for
helium-core white dwarfs of very  low mass (0.160 and 0.148 M$_\odot$)
with  metallicity  $Z=0.02$.  These computations  provide  theoretical
support  to search for  and identify  white dwarfs  of very  low mass,
specially  useful  for  recent  and future  observational  studies  of
globular cluster, where these objects have began to be detected.

\end{abstract}


\section{Introduction } \label{sec:intro}

It is  generally accepted that  helium-core white dwarf (He  WD) stars
constitute the  final product  of the evolution  of some  close binary
systems. This is  so because an isolated star  would need a time-scale
much longer  than the  present age of  the universe  to reach a  He WD
configuration. The mass  of these objects should be  smaller than that
required  for  degenerate helium  ignition,  $\approx 0.45$  M$_\odot$
(Mazzitelli 1989), thus  resulting in very low mass  WDs. Low mass WDs
have been detected in large  surveys (Bragaglia et al. 1990; Bergeron,
Saffer \&  Liebert 1992; Bragaglia, Renzini \&  Bergeron 1995; Saffer,
Livio \& Yungelson 1998) and  represent an appreciable fraction of the
total population of WD stars.

The interest in  the study of these stars  has greatly increased since
their  binary nature  was placed  on a  solid observational  ground by
Marsh (1995)  and Marsh,  Dhillon \& Duck  (1995). Indeed, He  WD have
been  detected in  numerous binary  systems containing  usually either
another WD  or a millisecond  pulsar (Moran, Marsh \&  Bragaglia 1997;
Orosz et  al. 1999;  van Kerkwijk et  al.  2000 amongst  others). Such
systems are very interesting because they enable us to infer the properties
of one component  of the system from studying  the physical properties
of  the  other. This  is  important  in  connection with  pulsar  mass
determinations and the structure and evolution of WD stars.

The recent  detection of low mass  He WDs (M $\la$  0.20 M$_\odot$) in
compact binaries  belonging to globular  clusters (Cool et  al.  1998;
Edmonds et al. 1999; Taylor et al. 2001; Edmonds et al. 2001) has also
sparked the  attention of many researchers.  The  interest in studying
He WDs in  globular cluster is motivated not  only by their importance
in the  understanding of  the formation and  evolution of  the compact
binaries in  which these stars are  found but also  by the possibility
they offer of constraining globular cluster dynamics and evolution.

Numerous studies have been devoted to the evolution of He WDs. Amongst
them  we mention  those  of  Benvenuto \&  Althaus  (1998), Driebe  et
al. (1998), Hansen \& Phinney  (1998), Sarna, Antipova \& Ergma (1999)
and Althaus, Serenelli \&  Benvenuto (2001). In particular, Althaus et
al. (2001)  have investigated the  effect of element diffusion  on the
evolution of He WDs taking  into account the evolutionary stages prior
to the WD formation. These authors find that element diffusion induces
thermonuclear  hydrogen shell  flashes in  He WD  models  with stellar
masses in the range $0.18  \la $ M/M$_\odot \la 0.41$.  In particular,
Althaus   et  al.    (2001)   show  that   the   occurrence  of   such
diffusion-induced flashes leads to He WD models with hydrogen envelope
masses too small to support any further nuclear burning, thus implying
much  shorter  cooling  ages  than  in  the  case  when  diffusion  is
neglected. As shown by Althaus  et al. (2001), such short cooling ages
remove the  discrepancy between the  spin-down age of  the millisecond
pulsars B1855+09,  PSR J0034-0534 and  PSR J1012+5307 and  the cooling
ages of  their He WD companions. Evolutionary  calculations of Althaus
et  al.   (2001) have  received  tentative  support  from the  optical
detection of the WD companion  to the millisecond pulsar in 47 Tucanae
(Edmonds et al. 2001).

Colours and magnitudes appropriate for old He WDs with the predictions
of stellar evolution and  element diffusion have recently been derived
in a self-consistent way by Serenelli et al. (2001). In the Serenelli et
al.  (2001)  calculations, emphasis  is placed on  the late  stages of
evolution where  WD cooling is  strongly affected by the  treatment of
the outer layers, particularly  of the atmosphere.  These authors have
explored at  some length the evolution  of He WDs and  found that when
the effective temperature decreases below 4000K, the emergent spectrum
of  these  stars becomes  bluer  within  time-scales of  astrophysical
interest. Because Serenelli et al.  (2001) were interested in the late
stages of He WD evolution, they did not attempt a detailed modeling of
the  emergent spectrum of  these stars  at high  effective temperature
stages ($T_{\rm eff} \ga 8000$ K).

In this paper we improve  the calculations of Serenelli et al. (2001)
by including  a more detailed  treatment of the  microphysics entering
the WD atmosphere. This enables us to derive accurate colours and
magnitudes for He WDs at high effective temperatures where the effects
of line  broadening opacities are  not negligible. Another aim  of the
present work  is to extend the evolutionary  calculations presented in
Serenelli et al. (2001) to less massive He WD models on the basis of a
more physically sound  treatment of the mass transfer  stage than that
attempted in Serenelli et al.(2001). This will allow us to compare the
predictions of our evolutionary models with the observations of the He
WD candidates recently reported in  the globular clusters NGC 6397 and
47  Tucanae  by Taylor  et  al.  (2001)  and  Edmonds  et al.  (2001),
respectively.

The magnitude calculations in  the ultraviolet and optical regions are
strongly   sensitive  to  the   inclusion  of   hydrogen-line  opacity
(particularly for $T_{\rm eff} \ga  8000$ K), which has been neglected
in our  previous model grids  (Rohrmann 2001, Serenelli et  al. 2001).
For the present work, we include this additional opacity source in the
atmosphere code  and we re-calculate  the models of Serenelli  et al.,
yielding a  better prediction  of the colour indices for  hot He
WDs.   Furthermore, the  evaluation of  non-ideal effects  in  the gas
allows us to calculate more realist atmosphere models for WDs at both
early and  advanced stages of  the cooling.  In fact,  WD atmospheres
are well known to be characterized by high densities (so high as $\rho
\approx  0.1$  gr/cm$^3$)   where  interactions  amongst  neighbouring
particles, charged and  neutral, affect the number of  bound states of
atoms  and  molecules. These  particle  perturbations  can modify  the
chemical  equilibrium of  the  gas and  its  optical properties.   The
occupation probability formalism of  Hummer \& Mihalas (1988) provides
a tool  to treat  non-ideal effects in  the gas of  stellar envelopes.
This  theory avoids  typical rough  truncations of  particle partition
functions giving a  thermodynamically self-consistent way to calculate
the ionization and excitation  equilibrium in the gas.  The occupation
formalism  was introduced  by Bergeron  (1988)  in the  context of  WD
atmospheres and we have chosen it for the present work.

The  application of  Hummer-Mihalas formalism  for  calculating atomic
populations  necessitates corresponding  improvements in  the formulae
used to  compute the optical properties of  the gas.  Phenomenological
proposals of  simulations of the  optical spectra in the  framework of
the occupational  formalism have been  given by D\"appen,  Anderson \&
Mihalas (1987)  and Hubeny, Hummer  \& Lanz (1994).   These researches
introduce   the   calculation   of  the   so-called   pseudo-continuum
absorption, which  rises from ionization of  atoms strongly perturbed.
This opacity source  becomes very important in WD  stars for different
ranges of  effective temperatures, particularly when the
Balmer lines contribute significantly to the total opacity
(e.g. Bergeron, Wesemael \& Fontaine 1991).  In view of the
preceding  considerations,  we found  it  necessary  to  adopt here  a
particular optical representation for our atmosphere calculations.  It
is worth noting  that the opacity theory for  non-ideal gases is still
in a preliminary stage and additional studies are necessary.

In addition to  the improvements perfomed on the  equation of state of
the gas,  new quantum mechanical data  of collision-induced absorption
(CIA) processes  have been incorporated in our  atmosphere code. These
new data  yield significant changes  in the emergent  radiative energy
obtained for cool He WD models as compared with previous calculations.

The paper is  organized as follows: \S 2  describes the new treatments
of the input physics in the  atmosphere code.  In \S 3, the atmosphere
models are  coupled with evolutionary calculations  and news, accurate
results  are given for  very low  mass He  WD stars.   Conclusions are
presented in \S 4.

\section{Model Atmospheres } \label{sec:atm}

The  model atmospheric  structures were  computed by  means of  a code
developed  recently (Rohrmann  2001), which  has been  improved  in a
number of  aspects.  In what follows,  we describe at  some length the
main improvements we have included in our code.

\subsection{Non-ideal effects in the gas equation of state } \label{sec:nid}

One  of  the  most  important  improvements considered  in  this  work
concerns the incorporation in our code of the occupational probability
formalism of Hummer  \& Mihalas (1988) for the  treatment of non-ideal
effects in the gas equation of  state.  This formalism is based on the
free-energy-minimization technique  (cf.  Graboske, Harwood  \& Rogers
1969) and can  be synthesized by  writing, for each  chemical species,
the LTE  (local thermodynamic  equilibrium) population of  an internal
state i ($n_i$) relative to  the total particle density ($n_{tot}$) of
this species as
\begin{eqnarray}  \label{popi}
 \frac {n_i} {n_{tot}} = \frac {w_i g_i exp(-E_i/kT)} Z \;,
\end{eqnarray}
where  $w_i$ is  the so-called  occupation probability  of  level $i$,
$g_i$  the statistical weight,  $E_i$ the  excitation energy,  $Z$ the
internal partition  function, $k$ the  Boltzmann constant and  $T$ the
gas temperature.

Factors $w_i$ are given by  partial derivation of a non-ideal term $f$
in the Helmholtz free energy of the gas,
\begin{eqnarray}  \label{wi}
 w_i= e^{ \left( - \partial f / \partial n_i \right) / kT } \;.
\end{eqnarray}
The probabilistic interpretation of $w_i$ allows Hummer and Mihalas to
combine  the action  of statistically  independent  interactions. They
have  considered two  types  of  perturbations.  One  of  them is  the
excluded  volume  interaction  (also  called perturbation  by  neutral
particles), which arises from the  finite size of particles with bound
electronic states.  These perturbations are taken into account through
the second virial coefficient in  the van der Waals equation of state,
which implies an occupation probability of the form
\begin{eqnarray}  \label{wi_n}
w_i(neutral)=\exp  \left[   -\frac  {4\pi}  3  \sum_{j}   n_j  (r_i  +
r_j)^3\right] \;,
\end{eqnarray}
where the sum runs over all species $j$ having bound levels, $r_i$ and
$r_j$ being the effective interaction radii of involved particles.

The  other type of  perturbation of  atomic levels  is due  to charged
species.   In the  Hummer-Mihalas formalism,  these  perturbations are
calculated  from  a  fit  to  a quantum  mechanical  Stark  ionization
theory.  This yields an  occupation probability  that can  be formally
expressed as
\begin{eqnarray}  \label{wi_c}
w_i(charged)= Q(\beta) \;,
\end{eqnarray}
which is  a fit over the  distribution of the  electric field strength
$\beta$ arising from the  charged particles. Originally, the Holtsmark
distribution   was    adopted   as   the    micro-field   distribution
function. However,  this distribution does  not take into  account the
correlations   of  the   charged  perturbers.   A   more  appropriated
distribution,  which includes  correlation effects,  was  presented by
Nayfonov  et al.  (1999).  The  numerical  fits given  by Nayfonov  et
al. to  compute $w_i(charged)$ have been adopted  in our calculations.
Finally, assuming  statistically independent perturbations,  the joint
occupation  probability   of  an  atomic  state  is   the  product  of
$w_i(neutral)$ and $w_i(charged)$.

In our calculations  the atmospheric gas is assumed  to be composed by
``bound''  species  (particles with  electronic  structure) H,  H$_2$,
H$^{-}$, H$_2^{+}$, H$_3^{+}$, He and He$^{+}$, and ``bare'' particles
H$^+$,  He$^{++}$ and e$^-$.   The energy  levels for  bound particles
have  been  chosen to  be  those of  the  isolated  atom or  molecule.
Eigenenergy  shifts  could  be  expected  as the  density  is  raised.
However, according to  experimental measurements (e.g. Weise, Kelleher
\& Paquette  1972), unperturbed energy levels is  a good approximation
at low densities  ($\log \rho \la -2$) and  seems reasonable at higher
densities (Saumon \& Chabrier 1991).

Characteristic  atomic and molecular  radii are  required in  order to
evaluate  the  excluded  volume  perturbations.  The  equivalent  hard
sphere radii (in \AA) for H$_2$ and H are computed from lineal fits to
results of Saumon \& Chabrier (1991, fig. 6),
\begin{eqnarray}  \label{rh}
 r_{H} = 1.876 - 0.294 \log T + (-0.576 + 0.100 \log T ) \rho \;,
\end{eqnarray}
\begin{eqnarray}  \label{rh2}
 r_{H_2} = 2.226 -0.331 \log T + (-0.975 + 0.175 \log T ) \rho \;.
\end{eqnarray}
which are based on  a temperature- and density-dependent thermodynamic
criterion (Weeks, Chandler, Andersen, 1971).  Radii for excited states
are obtained by simple scaling laws.

The    molecular    hydrogen    H$_2$    is   represented    by    193
vibration-rotational  states with  energy values  calculated  from the
expression  given  by  Herzberg  (1950), which  includes  anharmonicity
corrections,   deviations  from   rigid  rotator   approximation,  and
vibration-rotation coupling.   Excited electronic states  of H$_2$ are
above the  ground state by  more than  11 eV ,  so they can  be safely
neglected.   Effective  radii  for  vibration-rotational  states  were
calculated following  Mihalas et al. (1988), i.e.,  the equations (11)
and (12) of Vardya (1965) were solved for each state. These radii were
then scaled to ground radius (Eq. \ref{rh2}).

The neutral helium is represented by 144 ($n$, $L$, $S$) states ($1\le
n \le 50$), with energy values  for $n<22$ based on the data by Martin
(1973).  For  $n>22$, energies (in  eV) relative to ground  level were
estimated with the expression
\begin{eqnarray}  \label{eHe}
  E_n = 24.5876 - 1.002866 \frac {\chi_H} {n^2} \;,
\end{eqnarray}
where  $\chi_H$ is the  ionization potential  of the  atomic hydrogen.
Eq.  (\ref{eHe}) is  a  fit to  Martin's  data for  $15<n<23$ and  its
extrapolation to  $n>22$ has an  estimated error lower  than 10$^{-3}$
eV.  Effective radii  of the excited states of  He were evaluated from
the hydrogenic formula (performing  the calculation of the expectation
value of $r$ with the well-known electronic wave functions)
\begin{eqnarray}  \label{rHynl}
  \left<  r  \right>  _{(n,l)}  = \frac  {r_o}{3}  \left[  3n^2-l(l+1)
  \right] \;,
\end{eqnarray}
which  is  normalized  so that  the  ground  state  radius is  $r_o  =
3a_0/2Z_a$ ($a_0$ is the Bohr radius and $Z_a$ the atomic number). The
expression (\ref{rHynl}) yields a value  of 0.40 \AA \, for the radius
of both He  and He$^+$ ground states.  However, we  can expect that He
radius  is greater  than  He$^+$  radius due  to  its smaller  binding
energy. Therefore, we adopted for He ground state a radius of 0.50 \AA
\, as  used by  Mihalas et al.  (1988).  We  rescaled the radii  of He
excited states according to this $r_o$ value.

For the  atomic hydrogen we used  100 energy levels  calculated in the
standard  form,  and the  effective  radii  were  evaluated using  the
expression  (\ref{rHynl}) averaged over  the azimuthal  quantum number
$l$,
\begin{eqnarray}  \label{rhyn}
 \left< r \right> _{(n)} = \frac {r_o}{6} \left( 5n^2 + 1 \right) \;.
\end{eqnarray}
with $r_o$ given in  Eq. (\ref{rh}). Similarly, He$^+$ was represented
by   250   hydrogenic   energy   levels  having   radii   defined   by
Eq. (\ref{rhyn}) with $r_o = 3a_0/2Z_a = 0.40$ \AA.

Following Lenzuni  \& Saumon (1992), we have  chosen as characteristic
radii of H$^{-}$, H$_2^{+}$, H$_3^{+}$  the values 1.15, 1.32 and 0.65
\AA, respectively.  For each of these  species we assigned  an occupation
probability  corresponding to  a representative  state with  an energy
given  by the  ionization  or dissociation  energy  of the  considered
particle.

In the  frame of the  occupation probability formalism, the  ideal gas
equation  of  state  (EOS)  which relates  pressure,  temperature  and
density, must  be corrected taking  into account the non-zero  size of
the  molecules  and  atoms.   A simple  interpolation  formulae  which
satisfies this  requirement follows from the van  der Waals's equation
(e.g. Landau \& Lifshitz 1980, p. 232)
\begin{eqnarray}  \label{pnkt1}
 P_g = \frac {n_p k T} {(1-bn_p)} \;,
\end{eqnarray}
where $bn_p$ (with $n_p$ being  the numerical density of particles) is
proportional to the ratio of the volume occupied by extended particles
to the total available volume,
\begin{eqnarray}  \label{pnkt2}
 bn_p = a \frac {4\pi} 3 \sum_{j} r_j ^3 n_j \;.
\end{eqnarray}

\begin{figure}
\vskip 0.3cm
\begin{center}
{\epsfysize=170pt \leavevmode \epsfbox{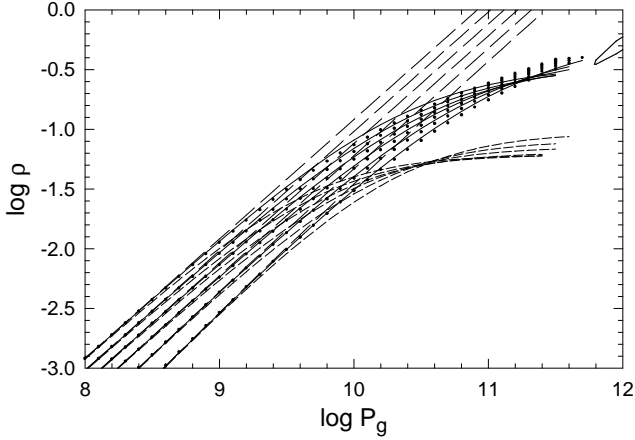}}
\end{center}
\caption{Comparison    of    density    isotherms   calculated    from
Eq.  (\ref{pnkt1}) with  $a=0$ (long  dashes), 1  (solid lines)  and 4
(short dashes).   Dots show  the EOS-SC predictions.   The temperature
decreases from  $\log T=3.8$  (right) to $\log  T=3.3$ (left),  with a
constant spacing  of $\Delta \log  T=0.1$.  The heavy line  denote the
plasma phase transition predicted by the EOS-SC (see text).}
\label{f:pgrho}
\end{figure}

A virial expansion at second order  of the EOS of a gas formed by
hard spheres results in $a=4$.  However, the value  of the  constant $a$
must be chosen  so as to give the best  agreement with experiments or,
otherwise, with the best  model predictions.  In our case, we have
chosen the equation of state of Saumon \& Chabrier (EOS-SC) to perform
comparisons with our gas model. EOS-SC  is based on a careful study of
inter-particle interactions for hydrogen and helium gases (see Saumon,
Chabrier \& Van Horn 1995).  As illustrated in Fig. \ref{f:pgrho}, a
value $a=1$ reproduces the EOS-SC results  at $\log
\rho \la  -0.6$, which is sufficient  for calculations of  He WD model
atmospheres   in   the   studied   range  ($T_{\rm   eff}\ge   2500$).
Calculations displayed in Fig.  \ref{f:pgrho} correspond to a hydrogen
gas with temperatures between 1995 and 6310 K.  The ideal gas equation
($a=0$)  predicts high  densities as  the pressure  increases, whereas
Eq. (\ref{pnkt1}) with $a=4$  overestimates the excluded volume effect
and yields too high a pressure for a given density.  At low density all
models recover  the perfect gas limit.   It is interesting  to note in
the figure  the location of  the plasma phase transition  predicted by
the EOS-SC, where the  pressure ionization is a discontinuous process.
The  portion  of  the  $\rho,  P$  plane bounded  by  the  heavy  line
corresponds to coexistent molecular  and metallic hydrogen states.  In
the right-hand boundary curve, starting from the critical point $(\log
P_g,\log  \rho)=(11.788,-0.456)$, lies the  phase dominated  by H$_2$,
whereas  the left-hand  boundary  curve corresponds  to the  partially
ionized phase dominated by fluid metallic hydrogen.

\begin{figure}
\vskip 0.3cm
\begin{center}
{\epsfysize=160pt \leavevmode \epsfbox{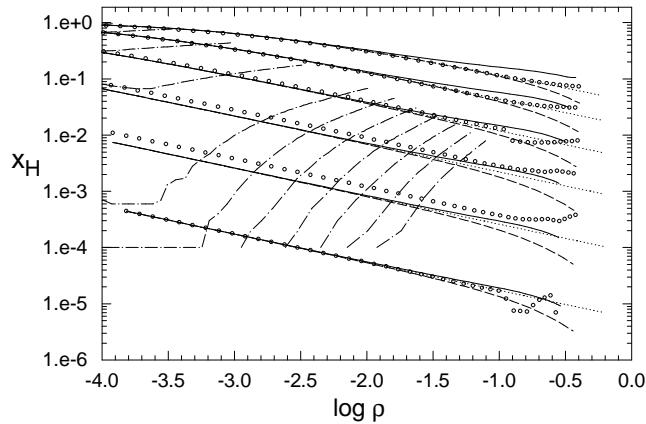}}
\end{center}
\caption{Concentration of atomic hydrogen along isotherms (from bottom
to top)  $\log T  = 3.3$, 3.4,  3.5, 3.6,  3.7 and 3.8.  Dotted lines,
dashed lines and circles  represent calculations using the ideal gas model
(with  a  density-dependent cutoff  in  the  partition  function of  H
atoms), occupational  probability formalism and  EOS-SC, respectively.
Solid  lines are calculated  with the  occupation formalism  using the
H$_2$ effective radius increased to 1.4 of the original value given by
Eq. \ref{rh2}.  Dash-dotted lines represent atmosphere models at $\log
g=8$ with  $T_{\rm eff} = 2500$  (250) 4000 (500) 6000  (from right to
left).}
\label{f:sau_eos}
\end{figure}

In connection with our implementation  of a non-ideal gas model for He
WD  atmosphere  calculations, some  words  are  in  order.  Saumon  \&
Chabrier   (1991)   found  that   non-linear   contributions  to   the
configuration  free energy  (based on  fluid perturbation  theory) are
necessary to yield the  pressure dissociation of H$_2$ molecules, such
as it  is expected in  a hydrogen gas  at high density  ($\rho \approx
0.5$  g/cm$^3$).   They  also  show  that  a  simple  excluded  volume
interaction  (such as  that  used in  our  gas model)  cannot lead  to
pressure ionization.   This result  is confirmed by  our calculations,
which are  based on  the non-ideal gas  model detailed in  the present
subsection.   We  display   in  Fig.  \ref{f:sau_eos}  the  fractional
abundance of atomic hydrogen along isotherms calculated with different
gas  models.  Results  based on  an ideal  gas, our  rigid  sphere gas
model, and the  EOS-SC, are represented by dotted  lines, dashed lines
and  open circles, respectively.   As density  grows above  $\log \rho
\approx  -2$, we  found  that a  non-ideal  contribution to  Helmholtz
energy based  {\em only} on the excluded  volume interaction increases
the molecular formation respect to ideal gas model calculations (using
classical  Saha relation).  On  similar conditions,  nevertheless, the
atomic  abundance   along  isotherms  computed   with  EOS-SC  becomes
relatively      shallow     before      reaching      the     pressure
ionization-dissociation  regime.   Therefore,  it seems  essential  to
include non-linear contributions  in the configuration energy (besides
the occupation  probability formalism)  of  the gas  for a  proper
treatment of  pressure dissociation. However, we found  it possible to
represent with  our model the gas behavior predicted by  the EOS-SC,
adjusting  the  effective  radius   of  molecules  H$_2$  (instead  of
implementing a  more sophisticated  gas model).  In  order to  obtain a
chemical equilibrium  at high density in better  agreement with EOS-SC
calculations, we found it appropriated to use a H$_2$ effective radius
increased to 1.4 of  its original (temperature- and density-dependent)
value given by Eq. (\ref{rh2}).
\footnote{The new values  of $r_{H_2}$ (1.1 - 1.5  \AA) applied in our
model  fall  within  the  dispersion  of  values  found  by  different
researches,  see \S \ref{sec:res1}  } This  modification allows  us to
obtain   more  appropriate   chemical  abundances   (solid   lines  in
Fig. \ref{f:sau_eos}) in the  regime of hight densities characteristic
of  deep layers  of  cool WD  model  atmospheres (dash-dotted  lines),
without altering the gas model predictions for lower densities.

\subsection{Opacity laws for a non-ideal gas} \label{sec:opa}

It is observed in DA  WD spectra (i.e. spectra showing hydrogen lines)
that the  high Balmer lines  strongly overlap on  the red side  of the
Balmer  jump, forming  a  pseudo-continuum. This  occurs similarly  in
other spectral  series and it gives  rise to a rather  smooth curve for
the opacity law instead of abrupt jumps in the serie limits.

In the  model atmospheres, the evaluation  of the opacity  at the line
merging region is often based  on a cutoff of high-lying atomic levels
and a shifting  of the continuum edge down to the  last line rising in
the  atomic model.  According to  D\"appen et  al. (1987  -- hereafter
DAM),  the  occupational formalism  of  Hummer  \&  Mihalas offers  an
alternative approximation to compute pseudo-continuum opacities.

Imagine  two atomic  states, one  lower energy  level $i$  and another
upper energy level $j$  with occupation probabilities $w_i$ and $w_j$,
respectively. Based on the probabilistic interpretation of the factors
$w$, DAM consider that a fraction of $w_j$ of the atoms have the bound
state $j$ unperturbed, and the complementary fraction $1-w_j$ have the
level $j$ ``dissolved'',  i.e., an atom in this  latter fraction finds
itself unbound and can be treated  as a free level. The first class of
states are simply  called `$j$ bound' and the  second ones are refered
as `$j$ dissolved'.  From  this classification, DAM derive that atomic
transitions  $i \rightarrow  j $  have a  probability $w_j/w_i$  to be
bound-bound transitions and a probability $1-w_j/w_i$ to be bound-free
transitions (contributing to the pseudo-continuum opacity).  Hubeny et
al. (1994 --  hereafter HHL) proposed a generalization  of the optical
simulation  of   DAM  to   non-LTE  conditions,  but   conserving  the
qualitative aspects of the original picture.  Transition rates derived
from these simulations are summarised in Table \ref{tbl-1}, together with
an early proposal due to Mihalas (1966).  We denote $\Lambda _{ij}$ as
the transition probability per second from $i$ to $j$ corresponding to
an isolated atom.
              
\begin{table}
\caption{Simulations for  the atomic transition rate  $i \rightarrow j
$.}
\label{tbl-1}
\begin{tabular}{lrr}
\hline Model  & $j$ bound &  $j$ dissolved \\ \hline  Mihalas (1966) &
$w_iw_jn_i\Lambda _{ij}$ & $w_i(1-w_j)n_i\Lambda _{ij}$ \\ D\"appen et
al.        (1987)        &        $(w_j/w_i)n_i\Lambda_{ij}$        &%
$(1-w_j/w_i)n_i\Lambda_{ij}$ \\ Hubeny  et al. (1994) & $w_jn_i\Lambda
_{ij}$  &  $(1-w_j)n_i\Lambda   _{ij}$  \\  Present  interpretation  &
$w_jn_i\Lambda _{ij}$ & --\quad\quad \\ \hline
\end{tabular}
\end{table}

We note, however, that the optical  simulations of DAM and HHL are not
completely self-consistent  with the Hummer-Mihalas  formalism because
the  dissolved states  have been  incorporated  in the  theory {\em  a
posteriori}, when  the stoichiometric equations  were just established
for particles  species having {\em  unperturbed} eigenenergies.  Thus,
for example,  if (as  it is  permitted in DAM  and HHL  simulations) a
fraction of atoms in certain  level $i$ can undergo transitions to the
bound state  $j$ and a complementary  fraction of atoms in  $i$ can be
ionized at the  same energy (dissolved state $j$),  then the level $i$
has not an unique energy value  (that of isolated atom) respect to the
continuum as it is assumed in the Hummer-Mihalas framework. Indeed, in
these  simulations,  the  population  equilibrium is  calculated  with
ionization    potentials    without    plasma-induced   shifts,    but
simultaneously, certain fraction of atoms  are permitted to be left in
dissolved  states  (i.e.,  lies   in  the  continuum)  in  transitions
involving  energies {\em  lower} than  the ionization  energies.  This
inconsistence suggests a revision of the opacity theory.

Our interpretation of the gas optical properties in the Hummer-Mihalas
formalism is as  follow. First, the dissolved states  are not possible
in the  gas model,  because its existence  implies a violation  to the
particle distribution  such as is dictated by  the used stoichiometric
equations of  ionization. Then, the classification  of atoms according
to state $j$ bound or  dissolved is illegitimate for the present state
of  the theory.   Second, the  factors $w_i$  must participate  in the
atomic transition  rates, since they  decide over the  availability of
the   atomic   states,   such   as   it  is   clearly   expressed   in
Eq. (\ref{popi}).

Thus, the  rate of  transition $i  \rightarrow j $  must be  zero when
$w_j=0$  (i.e., when $j$  is an  unavailable state)  and equal  to the
standard rate  $n_i\Lambda _{ij}$  when $w_j=1$ (atom  not perturbed).
Intermediate rate  values should be found  when $0<w_j<1$.  Therefore,
the  simplest heuristic representation  of the  transition rate  for a
non-ideal gas (characterized by the Hummer-Mihalas formalism) is
\begin{eqnarray}  \label{tato3}
n_i \Lambda _{ij} w_j \quad .
\end{eqnarray}
We can interpret $\Lambda  _{ij} w_j$ as the (conditional) probability
that the atom in the state $i$ makes a transition to state $j$ if this
is available.
\footnote{This conditional  probability was also proposed  by HHL (see
Table  \ref{tbl-1}), but  as  a  result of  the  decomposition of  the
transition rate $n_i\Lambda _{ij}$ into two terms, one referred to $j$
as  a bound  state and  the  another term  corresponding to  $j$ as  a
dissolved  state.}  The  occupation probability  $w_i$ of  the initial
state   in  the  transition   is  not   explicitly  present   in  rate
(\ref{tato3}),  because the  availability of  starting atoms  for this
process  is just  given by  the present  population $n_i$  (this  is a
difference with the proposal of Mihalas 1966, see Table \ref{tbl-1}).

Despite our  critique of the current optical  simulations based on
the occupational  formalism, we considered  that the proposals  of DAM
and HHL  are (into  the {\em state-of-the-art})  useful approximations
for the representation  of the opacity in the  line merging region, at
least  for Balmer,  Paschen and  other  series (except  for the  Lyman
series, see  below).  In fact, for  interpreting current high-accuracy
spectrophotometric  observations,  models  including  pseudo-continuum
opacity  are necessary.   In this  sense, the  phenomenological theory
introduced  by  DAM provides  an  approximation  to  the solution.  In
the absence of a  self-consistent theory, for the opacity  laws applied to
our calculations,  we take from  the DAM proposal the  contribution to
the pseudo-continuum cross-section, such as we explain inmediately.

In our model atmospheres,  the total opacity coefficient (corrected of
stimulated emissions) containing all contributions of the level $i$ is
calculated as
\begin{eqnarray}  \label{tato4}
 \chi _{ij,\nu} =  n_i \left[ \sum_j \alpha _{ij,\nu}  w_j + D_{i,\nu}
 \alpha _{ik,\nu} w_k \right] \left( 1-e^{-h\nu/kT}\right) .
\end{eqnarray}
The first term  in the square brackets corresponds  to the bound-bound
opacities  (with  cross-section   $\alpha  _{ij,\nu}$)  following  the
prescription  (\ref{tato3}).   The  second  term  in  Eq.(\ref{tato4})
represents the bound-free absorption in the same prescription (i.e.,
with the  corresponding cross-section $\alpha  _{ik,\nu}$ multipled by
the occupation  probability of  the final state,  for example,  in the
case of  H atoms  $w_k=1$) modified with  a factor  $D_{i,\nu}$, which
takes into account the pseudo-continuum contribution.

The   so-called    dissolved   fraction   $D_{i,\nu}$    extends   the
photoionization  cross-section  to  frequencies  below  the  frequency
threshold corresponding to an  isolated atom. DAM give for $D_{i,\nu}$
the expression
\begin{eqnarray}  \label{tato5}
 D_{i,\nu} = \frac {w_i - w_{n^*} } {w_i} \quad ,
\end{eqnarray}
where  $w_{n^*}$  is  the  occupation  probability  calculated  for  a
fictitious  (unless $n^*$  is  integer) state  with effective  quantum
number $n^*$ given by
\begin{eqnarray}  \label{tato6}
n^* = \left( \frac {1}{n_i^2} - \frac{h\nu}{\chi} \right)^{-1/2} \quad
,
\end{eqnarray}
where $\chi$  is the ionization  potential of the  considered species.
For imaginary $n^*$, we define $ D_{i,\nu}=1$ and the usual bound-free
opacity is recovered in Eq. (\ref{tato4}).

Conceptually, the  opacity law given in  Eq.(\ref{tato4}) is different
from  those proposed by  DAM and  HHL. However,  in practice, all
these  proposals  give  similar  results because  the  most  important
spectral series  (Lyman, Balmer and  Paschen) have low  lying starting
levels,  where the  occupation probabilities  are very  close  to unity
($w_i = 1$).

Finally, it is worth noting  that the absorption law (Eq. \ref{tato4})
computed  with  $D_{i,\nu}$  given  by  Eq.  (\ref{tato5})  yields  an
unphysical pseudo-continuum Lyman opacity for WD stars (Bergeron, Ruiz
\&  Leggett 1997).   This  theoretical  opacity is  so  large that  it
becomes  dominant over  the total  opacity, even  in the  infrared.  A
modified dilution factor $D_{i,\nu}$  for Lyman opacity was considered
by  Bergeron (2001)  in  order to  fit  the spectrum  of  the halo  WD
0346+246.   However,  this  ad  hoc  proposal is  based  on  arbitrary
physical  assumptions  and   needs  further  investigation.   We  have
prefered not  to include  this source of  opacity in  our calculations
until a more appropriate theory becomes available.

\subsection{Hydrogen line opacity } \label{sec:hlo}

Line  opacities of  the Lyman,  Balmer  and Paschen  series have  been
included in  the present  model atmosphere calculations.   However, in
this work we are not  interested in reproducing detailed observed line
profiles  but instead  in computing  the radiated  energy distribution
including line  blanketing. So,  theoretical line opacities  have been
incorporated  in the  model atmosphere  calculation in  an approximate
way.

\begin{figure}
\vskip 0.2cm
\begin{center}
{\epsfysize=290pt \leavevmode \epsfbox{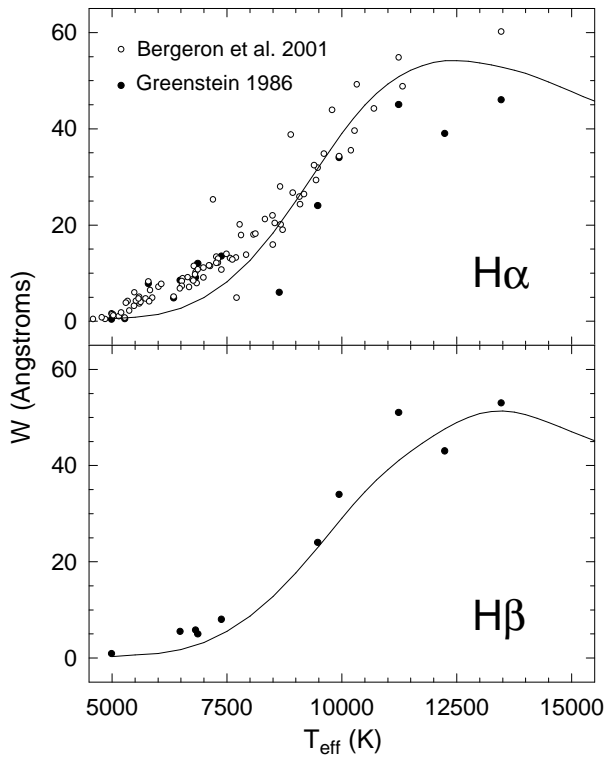}}
\end{center}
\caption{Equivalent widths of H$\alpha$ and H$\beta$ for a group of WD
stars,  as a  function of  $T_{\rm  eff}$.  Solid  lines indicate  the
predictions of our $\log g = 8$ models.}
\label{f:anchos}
\end{figure}

Because of the high densities encountered in WD atmospheres, the pressure
broadening is  usually the dominant line-broadening  mechanism and the
profiles ($\sigma_\nu$) can be assumed Lorentzian,
\begin{eqnarray}  \label{prof1}
 \sigma_\nu = \frac {\gamma} {\pi [ (\nu - \nu_o)^2 + \gamma^2 ] } \;,
\end{eqnarray}
where $\nu_o$ is the central frequency of the line.  The total damping
width $\gamma$  is the sum  of the natural,  linear Stark and  van der
Waals damping widths (in radians s$^{-1}$),
\begin{eqnarray}  \label{prof2}
\gamma = (\gamma_{nat} + \gamma_{Stark}/c + \gamma_{v d W}) / 4\pi \;.
\end{eqnarray}
The broadening  effects have  been considered with  damping parameters
given by Anderson  (1989). The linear Stark width is  a fit to results
of Griem  (1960), and the van  der Waals width  is a fit to  the $d-f$
transitions  of  Derrider  and  Van Rensbergen  (1976).   Taking  into
account the equivalent widths typically  observed for WDs, we found it
necessary to include  an additional parameter $c$ to  reduce the Stark
width given by the Anderson's formula (specially in hot stars, $T_{\rm
eff}  >10000$ K, where  the gas  is ionized  and the  Stark broadening
dominates  over  the  other   broadening  mechanisms). For that
purpose, we empirically found it appropriate to set $c$ equal to 3.
Fig.  \ref{f:anchos} shows  a
comparison  of the  equivalent widths  of the  H$\alpha$  and H$\beta$
lines observed for a group  of WDs (from Greenstein 1986 and Bergeron,
Leggett  \&   Ruiz  2001)  with   those  calculated  from   our  model
atmospheres.

Since  the radiative  flux  and atmosphere  structure is  particularly
sensitive to the profile of  L$\alpha$, a more accurate theory of line
broadening  was used  for  this line.  To  this end,  we employed  for
L$\alpha$  the  Stark-broadening  tables  of Vidal,  Cooper  \&  Smith
(1973).

In  hot model  atmospheres (e.g.  $T_{\rm eff}\approx  15000$  K), the
Lyman lines can  ``block'' up to 50 \% of  the energy transported in
the  continuum  spectrum.  Because   its  strong  influence  over  the
structure of the  atmosphere, Lyman line opacity has  been included in
the  energy  balance  and  transfer  equations  during  the  iterative
temperature correction procedure to compute each model atmosphere (see
Rohrmann 2001).   Then, from  the thermodynamic stratification  of the
converged  model,  we  computed   the  emergent  flux  including  also
formation of Balmer and Paschen  line series, the effect of which over
the temperature-pressure structure was considered negligible.

We  elected  about  six  hundred  frequency points  to  represent  the
radiation   field,   as   required   to  calculate   in   detail   the
emission-opacity  coefficients from all  continuum and  line processes
included in the atmosphere code.

\subsection{Collision-induced absorptions } \label{sec:CIA}

In  mixed hydrogen  and helium  gases,  colliding pairs  of atoms  and
molecules such as H$_2$-H$_2$, H$_2$-He and H-He can be sources of CIA
opacity.  Recent calculations  of H$_2$-He absorption coefficients are
available from  Jorgensen, Hammer,  Borysow \& Falkesgaard  (2000) and
cover  temperatures  between 1000  and  7000 K  for  the  25 --  20088
cm$^{-1}$ frequency  region.  Also, most up-to-date  CIA opacities due
to H$_2$-H$_2$ have been calculated by Borysow, Jorgensen \& Fu (2001)
for  $1000 \le  T \le  7000$  K and  frequencies between  20 --  20000
cm$^{-1}$.   Furthermore,  H-He  CIA  data  based  on  recent  quantum
mechanical computations  from Gustafsson  \& Frommhold (2001)  are now
available for temperatures 1500 -- 10000 K and frequencies 50 -- 11000
cm$^{-1}$.

\begin{figure}
\vskip 0.6cm
\begin{center}
{\epsfysize=320pt \leavevmode \epsfbox{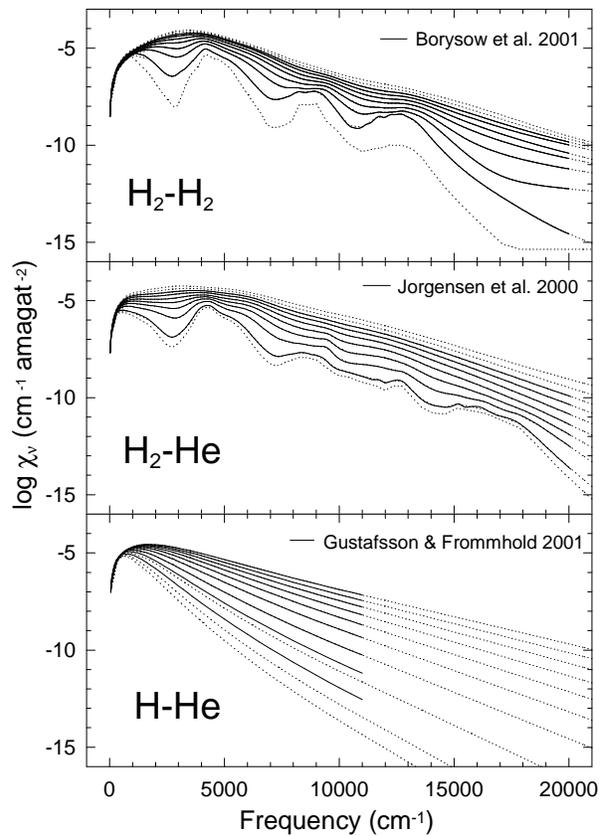}}
\end{center}
\caption{Collision-induced   opacity  corresponding   to  H$_2$-H$_2$,
H$_2$-He  and H-He  derived by  Borysow  et al.  (2001), Jorgensen  et
al. (2000)  and Gustafsson \& Frommhold  (2001), respectively.  Dotted
lines indicate our calculations based  on those results, for 500, 1000
(1000) 9000 K  from bottom to  top.  Solid lines show  calculations of
Borysow et al.  and Jorgensen et al. for  temperatures $T=1000$ (1000)
7000 K, and those of  Gustafsson \& Frommhold for $T=1500$, 2250, 3000
(1000) 9000 K (all opacities grow with increasing temperature).}
\label{f:cia} 
\end{figure}

All  these CIA  opacities  were included  in  our model  calculations.
Extrapolation (in  logarithmic scales) to temperatures and frequencies
beyond  the data  limits, when  necessary,  were used  in our  models.
Fig.  \ref{f:cia} illustrates the  CIA  values calculated  by
Borysow  and  co-workers  (solid  lines) and  our  corresponding  fits
(dotted lines).
      
\subsection{Photometric calibration of models } \label{sec:pca}

In  order to  compute colour  indices in  the  $UBVRI-JHKL$ photometry
system, a set of zero  points have been calculated.  These calibration
constants  are  necessary  to  convert the  theoretical  model  fluxes
$(F_X,F_Y)$   (integrated   over   filter   response   functions)   to
observational colours $X-Y$ (corresponding  to certain filters $X$ and
$Y$),
\begin{eqnarray}  \label{color}
 X-Y = -2.5 \log \left( F_X/F_Y \right) +C_{X-Y} \,.
\end{eqnarray}
The zero points  $C_{X-Y}$ were set by requiring that  the flux of the
star Vega has zero  colours. We  employed the  synthetic flux  of the
model ($T_{\rm  eff}=9400$ K,  $\log g =  3.95$) calculated  by Kurucz
(1979) for Vega,  and the  observed fluxes of  Vega measured  by Hayes
(1985).   The zero  points computed  from these  fluxes are  showed in
Table \ref{tbl-2},  together with the  constants given by  Bergeron et
al.  (1997), which we  used in  previous calculations  (Rohrmann 2001,
Serenelli et al. 2001).  For the present calculations of the broadband
colour indices  and for reasons  of internal consistency, we  used the
zero point  derived from  the Vega model  provided by  Kurucz, because
this flux covers all broadband regions.

\begin{table}
\caption{Zero  points  $C_{X-Y}$  computed  using  ($a$)  a  synthetic
spectrum  from Kurucz  (1979) and  ($b$) the  observed fluxes  of Vega
measured  by Hayes  (1985), and  those provided  by ($c$)  Bergeron et
al. (1997).}
\label{tbl-2}
\begin{tabular}{ccccc}
\hline Colour Index& $(a)$ & $(b)$  & $(c)$ \\ \hline $U-B$ & $-0.471$
& $-0.499$  & -  \\ $B-V$  & $+0.596$ &  $+0.625$ &  $+0.609$ \\  $V-R$ &
$+0.555$ & $+0.560$ & $+0.557$ \\ $V-K$ & $+4.839$ & - & $+4.886$ \\ $R-I$ &
$+0.703$ & $+0.700$ & $+0.704$ \\ $J-H$ & $+1.099$ & - & $+1.082$ \\ $H-K$ &
$+1.093$ & - & $+1.135$ \\ $K-L$ & $+1.854$ & - & - \\ \hline
\end{tabular}                                           
\end{table}

\section{Results} \label{sec:resul}

We have used the sets of  cooling sequences of Serenelli et al. (2001)
to determine accurate spectra and colour indices for He WD models with
stellar  masses 0.406,  0.360, 0.327,  0.292, 0.242,  0.196  and 0.169
M$_\odot$.  These models were  obtained using the WD evolutionary code
described in Althaus  et al. (2001) in a  self-consistent way with the
predictions  of detailed  non-gray model  atmospheres (Rohrmann 2001),
element diffusion and the history  of the WD progenitor.  More details
of our  current model grid  can be found  in the cited  literature and
references  therein.   Additionally,  two new  evolutionary  sequences
corresponding to 0.160 and 0.148  M$_\odot$ have been calculated for a
proper interpretation of recent observational data.

For  the present  work,  particular  attention has  been  paid to  the
calculation of the  synthetic spectra with an improved  version of the
atmosphere code,  which treats hydrogen line  blanketing and non-ideal
effects  in the  gas  within  the occupation  formalism  of Hummer  \&
Mihalas (1988), as described in \S 2.

\subsection{Consequences of the new input physics} \label{sec:res1}

We first analyse  the implications of using a  non-ideal gas model for
calculating  model  atmospheres  of   He  WDs.   The  consequences  of
non-ideal effects  in the  equation of state  for WD  atmospheres have
been  explored  previously  in  the context  of  hydrogen  atmospheres
(Bergeron Wesemael  \& Fontaine 1991;  Saumon \& Jacobson  1999), pure
helium models  (B\"ohm et al. 1977; Kapranidis  1983; Bergeron, Saumon
\&  Wesemael 1995)  and  models with  mixed  composition (Bergeron  et
al.  1995; Bergeron  2001).  Therefore,  we only  refer here  to those
results particularly significant for the present study of He WDs.

\begin{figure}
\vskip 0.3cm
\begin{center}
{\epsfysize=310pt \leavevmode \epsfbox{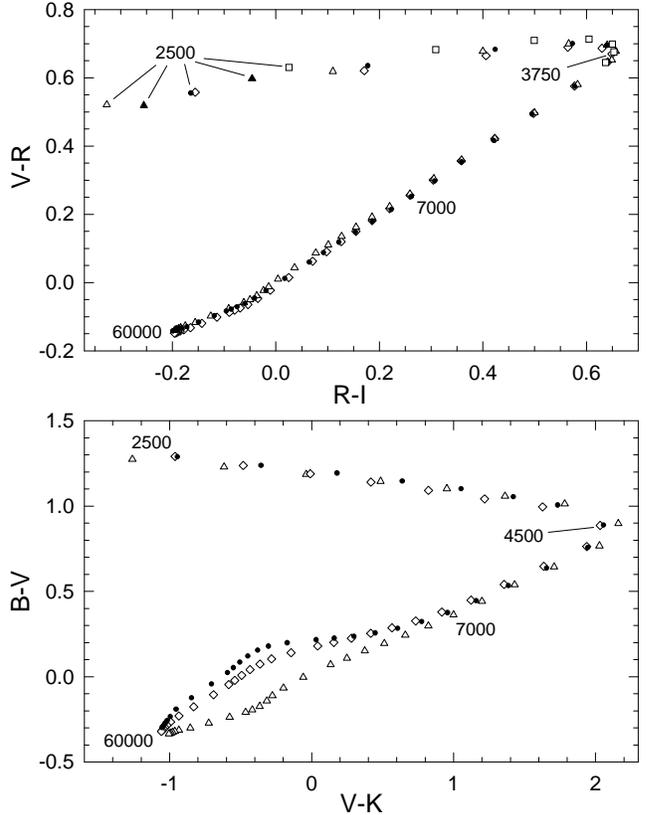}}
\end{center}
\caption{Colour-colour  diagrams for  $\log g=8$  models  with $T_{\rm
eff}=2500$  (250) 4000 (500)  10000 (1000)  17000, 20000  (5000) 60000
K.  Open  diamonds denote  our  current  calculations, whereas  filled
circles  and  open  squares  correspond  to  Bergeron  (2000,  private
communication) and  Saumon \&  Jacobson (1999, $T_{\rm  eff}\le 4000$)
models, respectively.  Our  old results based on ideal  gas theory and
without hydrogen-line and  pseudo-continuum opacities are indicated by
open triangles.  Some $T_{\rm eff}$  values are labelled on the plots.
The ($V-R$,  $R-I$) diagram shows also results  (filled triangles) for
the  $T_{\rm eff}=2500$ model  based on  non-ideal gas  with different
values of the effective radius of H$_2$ molecules (see text).}
\label{f:col_ber}
\end{figure}

We  found  that the  effects  of a  non-ideal  gas  is manifold.   The
interaction  amongst the  particles determines  through  the partition
functions the  chemical equilibrium of  the gas, which can  affect its
thermodynamic  behaviour (equations of  state and  response functions:
adiabatic  gradient, specific heats)  and opacity  properties (changes
operated  through  variations  of  the  abundances  of  absorbers  and
formation of pseudo-continuum spectra).  These modifications can yield
changes  over the  hydrostatic  stratification of  the atmosphere  and
emergent energy  spectrum of the He  WD stars in two  regimes: in cool
models characterized  by molecular formation  and in hot  models where
the  hydrogen  line  opacity  is  relevant.   We  can  appreciate  the
consequences of  this new input physics in  the colour-colour diagrams
displayed  in  Fig.   \ref{f:col_ber},   where  our  current  and  old
calculations  (Rohrmann   2001)  are  compared   with  those  obtained
precedently  from  other atmosphere  codes  (data  kindly provided  by
Bergeron and Saumon).

\begin{figure}
\vspace{0.0cm}
\begin{center}
{\epsfysize=240pt \leavevmode \epsfbox{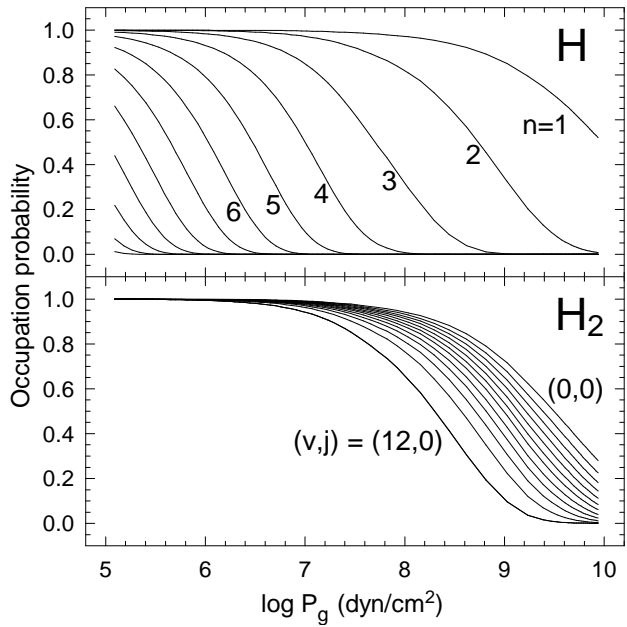}}
\end{center}
\caption{Occupation  probability  of  lower  levels  of  H  and  lower
vibrational  states $v$  of H$_2$  ($0\le v  \le 12$,  with rotational
quantum  number  $j=0$) as  a  function of  the  gas  pressure, for  a
0.292-M$_\odot$ He WD model with T$_{\rm eff}=2500$ K.}
\label{f:wwx}
\end{figure}

In cool atmospheres ($T_{\rm eff}  \la 4000$ K), which are composed of
a dense  fluid formed mainly by  H$_2$ molecules, the  main effects of
non-ideal gas have their origin in  a severe  reduction of  the occupation
probabilities  of  particle  states,  even  ground  states.   This  is
illustrated in Fig.  \ref{f:wwx}, where we note how low excitation
states of H and H$_2$  are perturbed with increasing pressure in the
surface of a 0.292-M$\odot$ He  WD at $T_{\rm eff}=2500$ K. Because of
the  low  temperatures  present  in  these  atmospheres,  the  charged
particle density  is low and  the main contribution to  the occupation
probabilities comes from the excluded volume interaction.  Molecules
have effective radii something higher  than the radius of the ground state
of the H atom and, therefore,  they experience perturbations  with larger
intensity.  However, high-lying levels of the hydrogen atom are weakly
bounded and  result easly destroyed, even  in the outer  layers of the
atmosphere.

It  is not possible  to predict  if an  increasing density  favours the
dissociation  or formation  of  molecules, because  this outcome  (the
chemical equilibrium) strongly depends on the radii assigned to atoms
and  molecules.    With  the  current  input   physics (involving  the
occupation  formalism  of  Hummer  \&  Mihalas),  we  found  that  the
non-ideal effects increase the H$_2$  dissociation into H atoms in the
atmospheric  region  where  the  continuum  radiation  is  formed  and
particularly   towards   deep   layers   in  cool   atmospheres   (see
Fig. \ref{f:sau_eos}).   As a consequence,  the importance of  the CIA
opacity on the radiative transfer  is reduced as compared with a model
based on ideal  gas. As a result, the resulting emergent  flux is less
displaced  to shorter  wavelengths respect  to our  early calculations
based on an ideal gas (Rohrmann  2001).  Fig. \ref{f:col_ber} shows for
$T_{\rm  eff}\le 3000$  K  some redder  colours  (specially $R-I$  and
$V-K$)  resulting  from  models  with non-ideal  gas  (open  diamonds)
compared  with   those  obtained   using  ideal  gas   theory  (open
triangles).  This  behaviour is in qualitative  agreement with results
reported by Saumon \&  Jacobson (1999) for hydrogen-model atmospheres.
Their  $\log  g  =8$ sequence  starting  at  $T_{\rm  eff} =  4000$  K
(reproduced on the $V-R$ versus $R-I$ diagram in Fig. \ref{f:col_ber})
appears  somewhat redder with respect to  our colour  computations, which
matches the sequence of Bergeron  very well.  A good agreement is also
obtained between our models and  those of Bergeron for the sequence in
($B-V$, $V-K$) diagram.

However, we note  that the effective radius assigned  to molecules can
strongly affect the spectrum of  very cool models. For example, if the
radius  of H$_2$  is reduced by 10\% of  the current  value used  in our
model, the formation of molecules is favoured, thus increasing the CIA
opacity and pushing the  emergent flux toward higher frequencies.  The
contrary effect follows when the  radius of H$_2$ is increased by 10\% of
the original value.  As a consequence, these exploratory modifications
yield, for example,  changes of about 0.1 $mag$ in  $R-I$ for cool WDs
(filled  triangles in Fig.  \ref{f:col_ber}).  Therefore,  we conclude
that a reliable description of the  spectrum of very cool WDs rests on
a satisfactory representation of molecular interactions.  However, the
investigation  of properties  of  dense molecular  hydrogen gas  often
yields different results according  to the techniques and methods used
(e.g.  Saumon, Chabrier  \&  Van  Horn 1995).   Thus,  for example,  a
variety of  effective radii have  been considered for  H$_2$ molecules
from different studies, e.g. $r_{H_2} =  1.48$ (Graboske, Harwood
\& Rogers 1969),  0.73 - 0.82 (Ross, Ree \&  Young 1983), 1.23 (Robnik
\& Kundt 1983),  1.45 (Mihalas, D\"appen \& Hummer  1987), 0.81 - 1.10
\AA (Saumon  \& Chabrier 1991).   Additional progress in the  study of
dense gases could prove useful in determining more accurate spectra for
very cool WDs.

For hot He WD atmosphere models  ($T_{\rm eff} \ga 8000$ K), the lower
atomic  levels  remain  almost   unperturbed,  so  that  the  chemical
equilibrium is not seriously  affected by non-ideal effects.  However,
perturbations amongst neighbouring particles  tend to destroy the high
energy levels  and affect  the line series  limits by  removing higher
series members.  In addition,  the continuum opacity advances over the
regions  where the  lines  overlap forming  a pseudo-continuum.   This
behaviour is described by the occupational formalism together with the
optical simulation synthesized in Eq. \ref{tato4}.

\begin{figure}
\vskip 0.2cm
\begin{center}
{\epsfysize=175pt \leavevmode \epsfbox{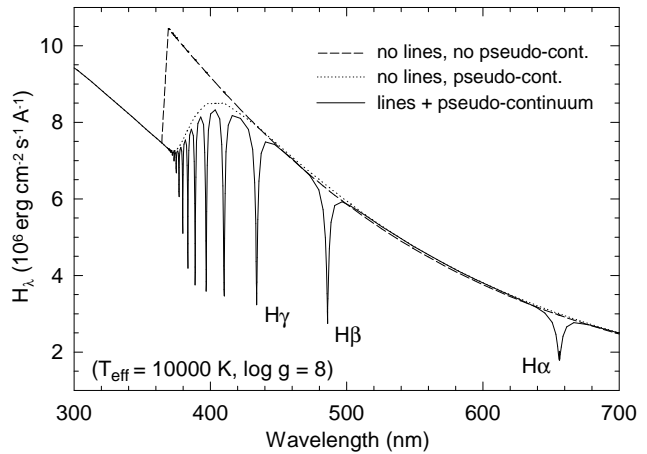}}
\end{center}
\caption{Emergent radiative  energy from  a hot H-model  atmosphere in
the   Balmer  line   region   using  different   opacity  laws.    The
pseudo-continuum  opacity simulated  with  the DAM  theory produces  a
shift  of the  Balmer discontinuity  towards the  red and  changes the
shape of the continuum spectrum.}
\label{f:balmer}
\end{figure}

We  analyse in  Fig. \ref{f:balmer}  the importance  of  the non-ideal
effects in the gas on the Balmer line spectrum.  Two spectra have been
calculated by neglecting the line opacities in order to illustrate the
importance of  the pseudo-continuum opacity.   For an ideal  gas model
without  line  absorption, the  energy  distribution  shows an  abrupt
Balmer discontinuity (dashed line in  the figure).  The inclusion of a
pseudo-continuum opacity  based on the DAM  optical simulation extends
the bound-free absorption toward  longer wavelength, yielding a marked
change  in  the  continuum  spectrum (dotted  line).   This  behaviour
results  from the  ionization of  perturbated H  atoms to  states with
energy lower that one corresponding  to isolated atom model.  When the
line opacity  is considered (solid line  in the figure),  we note that
the spectral  region where the lines overlap  is completetly dominated
by the pseudo-continuum opacity.  Accordingly, the intensity of higher
series  members  is  diminished  as  a  result  of  the  low  occupation
probabilities of final states in the bound-bound transitions.

\begin{table*}
\begin{minipage}{178mm}
 \label{tbl-bs}
\caption{Main properties of the atmosphere-codes whose results
are compared in Fig. \ref{f:col_ber}. Notation is as follow. 
DAM: optical simulation of D\"appen, Anderson \& Mihalas (1987); 
HM: occupation formalism of Hummer \& Mihalas (1988); 
NDHM: HM occupation formalism including plasma correlation effects from 
      Nayfonov et al. (1999); 
SCVH-EOS$^*$: equation of state (treating H$_2$, H, H$^+$ and e$^-$) based 
on the model of Saumon, Chabrier \& Van Horn (1995) and modified for
reproducing available high-pressure data (Saumon et al. 2000).}
\label{tbl-4}
\begin{tabular}{lllllll}
\hline 
Model atmosphere & $(P,T,\rho)$ relation & Chemical Equilibrium &
$r_{H}$ [\AA] & $r_{H_2}$ [\AA] & Opacity law & H$_2$-H$_2$ absorption \\
\hline  
Bergeron & Ideal gas & HM &  0.50$^{(a)}$ & - $^{(b)}$ & 
        DAM & Borysow et al. (1997) \\ 
Saumon \& Jacobson & SCVH-EOS$^*$ & SCVH-EOS$^*$ + HM($^c$)  & 0.65 & 1.07 &
        Standard & Borysow et al. (1997) \\ 
Present work & $Eq.(\ref{pnkt1})$ &  NDHM  & 0.58 - 0.86 & 1.1 - 1.5  & 
        $Eq.(\ref{tato4})$ & Borysow et al. (2001) \\ \hline
\end{tabular}
\end{minipage}

$^{(a)}$ Value adopted in order to reproduce observed profiles of higher Balmer lines 
(Bergeron et al. 1991).

$^{(b)}$ H$_2$ population was not computed in the HM formalism but 
with a standard Saha-like formula.

$^{(c)}$ Species H$^-$, H$_2^+$, H$_3^+$ are treated as perturbative species on 
the equilibrium using HM formalism (Lenzuni \& Saumon 1992).
\end{table*}

The colour calculations,  particularly those involving magnitudes from
broadbands lying over the Balmer  jump, are sensitive to the inclusion
of hydrogen line blanketing and pseudo-continuum process, as it can be
observed in  Fig. \ref{f:col_ber}.  Line opacity  and Stark ionization
strongly  contribute  to the  absorption  in  ultraviolet and  visible
spectra,  and significantly affect  the colours  for $T_{\rm  eff} \ga
8000$  K,  mainly  through  the  $B$ photometric  bandpass.   Our  new
calculations  agree  reasonably  well  with those  of  Bergeron.   The
remnant  discrepances amongst colour  calculations from  the different
atmosphere codes  shown in Fig.  \ref{f:col_ber}, could be due  to the
CIA  opacity used  for the  cool  end of  the model  sequence (we  use
recently available data) and,  presumably, to the constitutive physics
concerning the non-ideal effects used in the codes. 
Table 3 summarises the properties of the atmosphere codes,
the results of which are compared here.

\begin{figure}
\vskip 0.0cm
\begin{center}
{\epsfysize=230pt \leavevmode \epsfbox{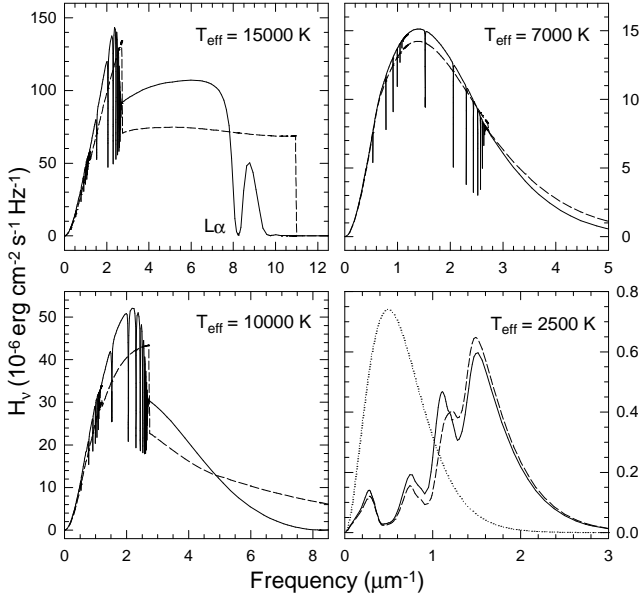}}
\end{center}
\caption{Spectra  calculated  for   different  $T_{\rm  eff}$  of  the
0.292-M$_\odot$  He WD  model. The  solid lines  indicate  the present
calculations   including  line  opacity,   non-ideal  gas   model  and
up-to-date  CIA opacity,  while  the dashed  lines  represent our  old
calculations without  these improvements.  For  $T_{\rm eff}=2500$, it
is  also displayed  the  blackbody spectrum  at  the same  temperature
(dots).}
\label{f:fm292}
\end{figure}

Some   representative   emergent  spectra   obtained   from  our   new
calculations   can  be   appreciated  in   the  cooling   sequence  of
0.292-M$_\odot$ He  WD model  displayed in Fig.  \ref{f:fm292}.  These
models correspond to $T_{\rm eff}=15000$, 10000, 7000, 2500 K, $\log g
= 7.092$, 7.195, 7.267, 7.384,  respectively, and have a pure hydrogen
atmosphere.  Several changes can  be appreciated in Fig. \ref{f:fm292}
comparing  the  new  spectra   (in  solid  lines)  with  our  previous
determinations (dashed lines) obtained  without line opacity and using
ideal gas theory. In hot models, the Lyman line series forces the flux
to emerge at  lower frequencies; even the global  change in the energy
distribution  of the  relatively cool  $T_{\rm eff}=7000$  K  model is
affected by  to Lyman line absorption  (due mainly to the  red wing of
L$\alpha$). As mentioned above, the  variations in the spectrum of the
cool model  ($T_{\rm eff}=2500$ K)  are due to both  non-ideal effects
and  updated CIA  opacity incorporated  in the  atmosphere  code.  The
difference with the blackbody emission (dots) is a remarkable property
of the advanced stages of cooling of He WD stars and WDs in general.

\subsection{Helium-core white dwarf photometry} \label{sec:res2}

\begin{figure*}
\begin{minipage}{178mm}
\vskip 0.7cm
\begin{center}
{\epsfysize=350pt \leavevmode \epsfbox{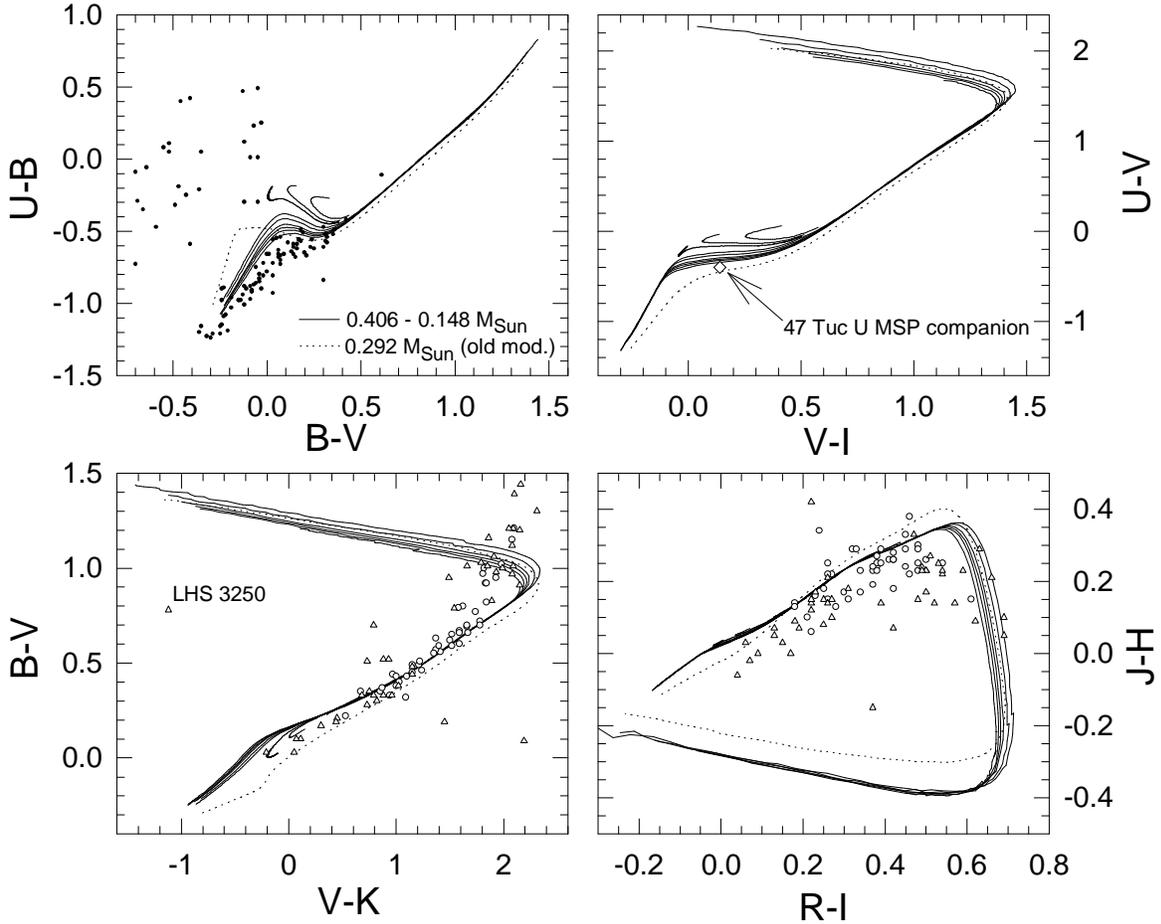}}
\end{center}
\caption{Colour-colour diagrams for  sequences corresponding to our He
WDs  with stellar  masses 0.406,  0.360, 0.327,  0.292,  0.242, 0.196,
0.169  0.160 and  0.148 M$_\odot$  (solid lines).  For  comparison, we
display the old 0.292  M$_\odot$ model sequence (dotted line).  Filled
circles  in the  ($U-B$,  $B-V$)  diagram correspond  to  a sample  of
seemingly  single DA  (hydrogen line)  and DB  (helium line)  WDs from
Saffer et al.  (1998).  The lower diagrams contain  observations of DA
(open  circles)  and non-DA  (open  triangles)  WDs  from Bergeron  et
al. (2001).  The  47 Tuc U MSP companion (denoted  by `U') (Edmonds et
al. 2001) and  LHS 3250 (Harris et al. 1999) are  shown in the ($U-V$,
$V-I$) and ($B-V$, $V-K$) diagrams, respectively.}
\label{f:col_col}
\end{minipage}
\end{figure*}

In  Fig.\ref{f:col_col} we show  some computed  colour-colour diagrams
for  our He  WD  models  (solid lines),  compared  to WD  observations
(symbols).  The cooling sequences start near the lower left corner and
extend towards  the upper  right corner before  showing a  turn-off in
some colours  such as  $V-I$, $V-K$ and  $J-H$.  The models  cover the
range  0.148-0.406 M$_\odot$.  In  the  case of  the lowest  mass
models (0.148, 0.160 and 0.169 M$_\odot$), the evolution after the end
of mass loss episodes is slow, so  they can age up to 1-1.5 Gyr before
reaching the knee  (highest $T_{\rm eff}$ point) and  starting to cool
down. In view  of this fact, there is  a loop at the hot  end of their
sequences    that   can    be   appreciated    in   the    curves   in
Fig. \ref{f:col_col}.   We notice that  our sequences fall  along the
bulk of  the observations and  expand towards cooler  regions ($T_{\rm
eff}\la 3500$ K) of the  diagrams, where there is little observational
information about  WD stars yet.  Dotted line  in Fig. \ref{f:col_col}
represents  our  old  computations   for  the  0.292  M$_\odot$  model
(Serenelli et al. 2001).  The differences between the new calculations
and the  previous approximations  are significant for  various colours
(affected   by   hydrogen-line   opacity   and   non-ideal   effects),
particularly at high and very low temperatures.

\begin{figure}
\vskip 0.2cm
\begin{center}
{\epsfysize=240pt \leavevmode \epsfbox{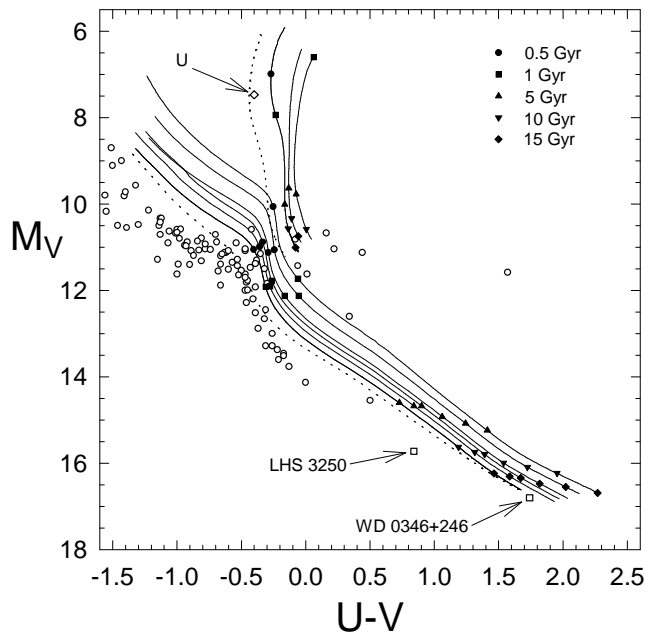}}
\end{center}
\caption{The  ($M_{\rm V}$,  $U-V$) magnitude-colour diagram. From the
top-right  corner, solid  lines  correspond to  He  WD sequences  with
0.148,  0.160, 0.169,  0.196,  0.242, 0.292,  0.327,  0.360 and  0.406
M$_\odot$. Dotted  lines display our  old models with 0.169  and 0.406
M$_\odot$.   Open circles  denote  a sample  of  apparently single  DA
(hydrogen line)  and DB (helium line)  WDs from Saffer  et al. (1998).
The 47 Tuc U MSP companion  (denoted by `U') and the ultracool WDs LHS
3250 and WD 0346+246 are also displayed on the plot.}
\label{f:mv_uv}
\end{figure}

\begin{figure}
\vskip 0.2cm
\begin{center}
{\epsfysize=240pt \leavevmode \epsfbox{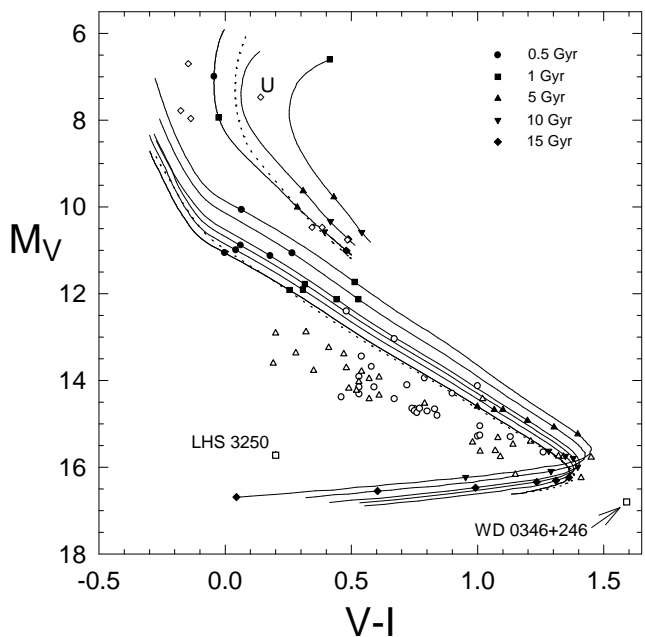}}
\end{center}
\caption{Same as Fig. \ref{f:mv_uv} but for the colour index $V-I$.  A
sample of  cool DA (open circles)  and non-DA (open  triangles) WDs is
included in the  figure. Open diamonds show canditate  He WDs observed
by Taylor  et al. (2001)  and the 47  Tuc U MSP companion  (denoted by
`U').}
\label{f:mv_vi}
\end{figure}

Figs.   \ref{f:mv_uv}  and   \ref{f:mv_vi}  display   absolute  visual
magnitude $M_{\rm V}$ versus $U-V$  and $V-I$ colours, respectively, for our
He WD models covering  0.148-0.406 M$_\odot$.  As already mentioned in
Serenelli et al. (2001), He WD models with mass greater than $\approx$
0.18 M$_\odot$ (in  the case of progenitors of  solar metallicity) can
reach high  magnitudes within cooling times  less than 15  Gyr, due to
thermonuclear  hydrogen  shell flashes  induced  by element  diffusion
(Althaus et  al. 2001).  Collision-induced absorption  by H$_2$ causes
turnover to bluer  $V-I$ at $M_{\rm V} \approx$ 15.5-16.5,  depending on the
stellar  mass, whereas  the  $U-V$ index  monotonically reddenes  with
decresing $T_{\rm eff}$. Molecular  opacity also yields a slow cooling
rate  in the  later  stages of  the  evolution.  In  the interests  of
comparison,  the old  computed tracks  for 0.169  and  0.406 M$_\odot$
(dotted lines in the  colour-magnitude diagrams) have been included in
the figures. Note that the current improvements in the atmosphere code
(especially the inclusion  of hydrogen-line opacity) yield significant
changes in the  colour values at hot WD stages  and, therefore, at low
magnitudes. The 0.169 M$_\odot$ sequence now appears approximately 0.2
$mag$ redder in the $U-V$ index, and 0.1 $mag$ bluer in $V-I$ at early
stages.  The open circles in  Fig. \ref{f:mv_uv} correspond to a group
of seemingly single WDs observed by Saffer et al. (1998), whereas open
triangles and  circles in Fig.  \ref{f:mv_vi} represent DA  and non-DA
WDs, respectively, from  Bergeron et al. (2001). It  is interesting to
note that our massive He WD  sequences follow the red edge of the bulk
of  the  observational  samples.   As  reference,  two  ultracool  WDs
($T_{\rm  eff}$  below 4000  K)  are  included  in the  figures  (open
squares), LHS 3250, characterized by a strong infrared flux deficiency
(Harris  et al. 1999),  and WD  0346+246, which  has an  atmosphere of
mixed  hydrogen  and  helium  composition (Oppenheimer  et  al.  2001;
Bergeron 2001).

As stated in  the introduction, very low-mass He WDs  have begun to be
detected  or  inferred  in  globular clusters.   Recently,  Taylor  et
al. (2001) have presented  observational evidence for He WD candidates
in the cluster NGC 6397.  Additionally, Edmonds et al. (2001) reported
the optical  detection of  the He  WD companion of  mass M  $\la$ 0.17
M$_\odot$ to a millisecond pulsar  in the globular cluster 47 Tucanae.
In  view of  these considerations,  we judge  it to  be  worthwhile to
extend  the evolutionary  calculations presented  in Serenelli  et al.
(2001) to less massive He WDs, thus enabling an appropriate comparison
with the  above-mentioned observations. To  this end, we  computed the
evolution  of  He   WDs  with  stellar  masses  of   0.148  and  0.160
M$_\odot$. Element  diffusion, nuclear burning and the  history of the
WD progenitor  are taken into account  (see Serenelli et  al. 2001 for
details). Realistic  initial models  are obtained by  abstracting mass
from a  1 M$_\odot$ model.  Because, in the  case of very low  mass He
WDs,  the  evolution  following  the  end of  mass  transfer  episodes
markedly depends  upon the  details of how  the envelope was  lost, we
have  attempted  here a  more  physically  sound  treatment to  obtain
realistic  initial He  WD models  than  that assumed  in Serenelli  et
al. (2001). Specifically,  the WD progenitor was supposed  to be in a
binary  system with  a 1.4  M$_\odot$ companion  (representative  of a
neutron  star)  and  during  pre-WD  evolution mass  loss  rates  were
adjusted so as to keep the radius of the progenitor close to the Roche
lobe  radius.   In order  to  compute  the  mass loss  rates,  angular
momentum losses  due to magnetic braking,  gravitational wave emission
and mass  loss were taken into  account (see Sarna,  Ergma \& Antipova
2000).   In  broad outline,  the  evolution  of  the 0.148  and  0.160
M$_\odot$ He WD models is  qualitatively similar to that of the lowest
stellar mass model analysed in  Serenelli et al.  (2001). Indeed, they
are  characterized by  the absence  of hydrogen  thermonuclear flashes
even  when  element diffusion  is  allowed  to  operate. The  lack  of
thermonuclear flashes forces the evolution of the star to be dominated
by stable nuclear  burning over most of the  entire evolution of these
models.   As a  result, they  are characterized  by very  long cooling
ages. Another resemblance with the  lowest mass models of Serenelli et
al.  (2001)  is the  fact that although  diffusion proceeds  much more
slowly than  in more massive models,  the outer layers  become  pure
hydrogen  on time-scales shorter  than evolutionary  time-scales.  The
evolution of these very low  mass He WD models in the colour-magnitude
diagrams is depicted in Figs. \ref{f:mv_uv} and \ref{f:mv_vi}.

Observational  data of  the He  WD star  detected, as  companion  to a
millisecond  pulsar (MSP),  in  47 Tucanae  by  Edmonds et  al. (2001)  is
included in Figs.~\ref{f:mv_uv} and  \ref{f:mv_vi}, where this star is
denoted by 'U'.  From the ($M_{\rm V},U-V$)  diagram we infer a mass slightly
higher than 0.17~M$_\odot$ for  this star, and from ($M_{\rm V},V-I$) diagram
a value somewhat lower than 0.16~M$_\odot$.  The pulsar characteristic
age of 2  Gyr (see Edmonds et al.  2001  and references cited therein)
is  consistent with the  age derived  for the  WD from  our theretical
models,  particularly if  the low  mass value  (0.16~M$_\odot$), which
yield 2-2.5 Gyr, is adopted. A lower value for the age is obtained for
a  0.17~M$_\odot$  model, resulting  in  a  WD  younger than  its  MSP
companion.  Evolution time-scales for these  low mass He WDs which do
not suffer from thermonuclear  flashes (Althaus et al. 2001) however,
are dominated  by hydrogen burning and  in this regard  we must recall
that   metallicity  of  the  WD progenitor  can  play an  important
role. He  WD models  presented in this  work were derived  from $Z=0.02$
progenitors and, given that $[{\rm Fe/H}]=-0.76$ for 47 Tuc, caution must be
taken when  trying to infer  the age for  this He WD with  the present
models.

In Fig.~\ref{f:mv_vi} we  also show the six He  WD candidates found in
NGC 6397 by Taylor et al. (2001). Absolute magnitudes were obtained by
using $(m-M)_{\rm V}=12.29$ and $V-I$ was dereddened  by using
E($B-V)=0.18$ and extinction laws for WFPC2 filters given by Holtzman  et
al. (1995).   The three brightest  stars of the observed  sequence are
located between  our 0.169 and 0.196  M$_\odot$ evolutionary sequences
in   Fig.~\ref{f:mv_vi}   suggesting  that   these   stars  could   be
$0.18-0.19$~M$_\odot$ He  WDs, whilst the  three dimmest ones  of that
sequence are compatible  with He WDs with stellar  masses between 0.16
and 0.17  M$_\odot$.  Another interesting aspect of  our sequences are
the  ages involved.   According to  our predictions,  the age  for the
three brightest  stars would be $\sim$  1 Gyr.  On the  other hand the
dimmest stars  seem to be  very old objects,  older than 8 Gyr  in all
cases and  even 10 Gyr for the  dimmest one. NGC~6397 is  a very metal
poor globular cluster $[{\rm Fe/H}=-1.95$], thus models with a much lower
metallicity  would  help give a  more  detailed  understanding  of  the
observations, since both mass and age determinations could be affected  
(e.g.,   the  mass   threshold  for  the   ocurrence  of
thermonuclear  flashes  is  a  function  of  the  metallicity  of  the
progenitor thus  affecting age determinations  even if the WD  mass is
known). In this  regard work is in progress to provide  a full grid of
He  WD models  for different  metallicities (Serenelli  et  al.  2002).
Recently, Townsley  \& Bildsten (2002)  have suggested that  the three
dimmest He  WD candidates reported  by Taylor et  al. could be  CO WDs
that accrete matter  from a 0.15 M$_\odot$ main  sequence companion at
an average rate of $10^9-10^{10}$ M$_\odot {\rm yr}^{-1}$. Additional
observations will be necessary to disentangle this issue.

\section{Conclusions} \label{sec:concl}

We   have  presented  new   computed  spectra,   colour-magnitude  and
colour-colour  diagrams  for cooling  sequences  of helium-core  white
dwarf  (WD)  stars.  These  models  include more detailed  physics
entering the atmosphere calculations. Non-ideal effects were carefully
taken into account in the computation of the chemical equilibrium of mixed
hydrogen and helium gases, using the occupation formalism of Hummer \&
Mihalas (1988).   We also included  a modified version of  the optical
simulation of D\"appen et al. (1987) to compute opacities in non-ideal
gases.   From these improvements  and the  evaluation of  line opacity
from  Lyman,  Balmer and  Pashen  series,  we  determine more  realist
atmosphere models of  He WDs at both early and  advanced stages of the
cooling.

Results  from our  earlier He  WD  calculations are  confirmed, but  a
detailed comparison  reveals some important  quantitative differences.
A clear advantage of the  newer treatment is the improved behaviour of
the emergent flux, especially in hot WDs ($T_{\rm eff}>8000$ K), where
the  line  blanketing   and  Balmer  pseudo-continuum  process  yields
substantial differences in ultraviolet and optical regions as compared
with our previous calculations, affecting colours in $UBV$ photometry.
The  use of  a non-ideal  gas model  and  up-to-date collision-induced
opacities  has also  allowed a  more reliable  calculation of  cool WD
spectra ($T_{\rm eff}<4000$ K).

We extended  our He WD model  grid by computing  cooling sequences for
stellar masses of 0.148 and 0.160 M$_\odot$. Comparison of the colours
evaluated for  our sequence of  models with recent  observational data
from He  WD candidates in the  cluster NGC 6397 reported  by Taylor et
al. (2001),  and the  He WD  companion to a  millisecond pulsar  in 47
Tucanae observed by Edmonds et al.   (2001), allow us to infer low mass
values for these  objets (ranging from 0.16 up  to 0.19 M$_\odot$). In
the  case of  47 Tuc,  our age  prediction for  the He  WD ($1-2.5$~Gyr)
agrees  reasonably well with  the pulsar  characteristic age  (2 Gyr).
However, given the low metalliticy of these clusters, a grid of He WD
models with low metallicity progenitors is needed in order
to contrast these and future observations with model predictions.
 
With  the present extensive  and physically  sound grid  of He  WD, we
provide a better  understanding of the spectral evolution  of low mass
WDs.  Our resulting He WD  models are particularly useful for interpretive
data  of very  low-mass WDs  within the  current  detectability limit,
allowing   to  determine   their  physical   parameters   (mass,  age,
temperature),  and  to  constrain   those  of  their  possible  binary
companions.

Detailed tabulations  of the results  presented here are  available at
http://www.fcaglp.unlp.edu.ar/$\sim$serenell  or upon  request  to the
authors at their email addresses.

Acknowledgements. 

We are grateful to P. Bergeron and D. Saumon for
informative collaborations on Table 3.
We also thank M. Barstow, our referee, for useful 
comments that improved the original version of our paper.

\begin{table*}
\begin{minipage}{177mm}
\caption{Selected stages for 0.148 and 0.160 M$_\odot$ He WD models}
 \label{tbl-3}
 \begin{tabular}{@{}cccrrrrrrrrrrr@{}}
\hline
$M_*/{\rm M_{\odot}}$ &
Log $T_{\rm eff}$ &  $Log (g)$ &  $Age$ (Myr) & U-B $\,$ & B-V $\,$ & V-R $\,$ &
V-K $\,$ & R-I $\,$ & J-H $\,$ & H-K $\,$ & K-L $\,$ & BC $\,$ & M$_{\rm V}$
$\,$ \\
\hline
0.148 & 3.8678 & 4.7665 &  998.9500 & -0.270 & 0.332 & 0.206 & 0.694 & 0.208 & 0.154 & -0.044 & 0.040 & -0.040 & 6.601 \\
   '' & 3.8916 & 4.9537 & 1329.7030 & -0.262 & 0.283 & 0.169 & 0.534 & 0.170 & 0.128 & -0.059 & 0.023 & -0.048 & 6.838 \\
   '' & 3.9058 & 5.1017 & 1602.4290 & -0.261 & 0.256 & 0.146 & 0.442 & 0.148 & 0.114 & -0.067 & 0.014 & -0.057 & 7.076 \\
   '' & 3.9138 & 5.2253 & 1848.7500 & -0.266 & 0.242 & 0.135 & 0.394 & 0.137 & 0.106 & -0.072 & 0.009 & -0.066 & 7.314 \\
   '' & 3.9189 & 5.3361 & 2083.1340 & -0.273 & 0.234 & 0.128 & 0.365 & 0.130 & 0.102 & -0.074 & 0.006 & -0.074 & 7.548 \\
   '' & 3.9217 & 5.4377 & 2317.4140 & -0.284 & 0.232 & 0.126 & 0.353 & 0.128 & 0.100 & -0.075 & 0.004 & -0.081 & 7.781 \\
   '' & 3.9222 & 5.5301 & 2557.7780 & -0.296 & 0.233 & 0.127 & 0.355 & 0.129 & 0.101 & -0.075 & 0.004 & -0.086 & 8.012 \\
   '' & 3.9205 & 5.6140 & 2809.5370 & -0.310 & 0.239 & 0.132 & 0.372 & 0.134 & 0.104 & -0.074 & 0.004 & -0.089 & 8.242 \\
   '' & 3.9171 & 5.6905 & 3073.4010 & -0.325 & 0.246 & 0.139 & 0.399 & 0.141 & 0.108 & -0.072 & 0.004 & -0.091 & 8.469 \\
   '' & 3.9120 & 5.7610 & 3352.8970 & -0.341 & 0.257 & 0.148 & 0.436 & 0.150 & 0.114 & -0.069 & 0.006 & -0.091 & 8.695 \\
   '' & 3.9057 & 5.8259 & 3648.9620 & -0.357 & 0.269 & 0.159 & 0.481 & 0.162 & 0.122 & -0.065 & 0.008 & -0.090 & 8.920 \\
   '' & 3.8980 & 5.8856 & 3963.8340 & -0.374 & 0.284 & 0.172 & 0.534 & 0.174 & 0.131 & -0.061 & 0.011 & -0.088 & 9.145 \\
   '' & 3.8891 & 5.9409 & 4304.4200 & -0.388 & 0.300 & 0.186 & 0.596 & 0.188 & 0.142 & -0.056 & 0.015 & -0.087 & 9.370 \\
   '' & 3.8792 & 5.9919 & 4677.4040 & -0.400 & 0.318 & 0.201 & 0.663 & 0.204 & 0.153 & -0.050 & 0.020 & -0.086 & 9.596 \\
   '' & 3.8686 & 6.0395 & 5102.4300 & -0.408 & 0.338 & 0.217 & 0.735 & 0.220 & 0.166 & -0.043 & 0.026 & -0.086 & 9.821 \\
   '' & 3.8502 & 6.1170 & 6128.4100 & -0.411 & 0.373 & 0.244 & 0.860 & 0.247 & 0.188 & -0.031 & 0.037 & -0.085 & 10.198 \\
   '' & 3.8404 & 6.1684 & 7643.8730 & -0.408 & 0.393 & 0.259 & 0.926 & 0.261 & 0.200 & -0.025 & 0.043 & -0.086 & 10.426 \\
   '' & 3.8297 & 6.2165 & 11150.6700 & -0.399 & 0.415 & 0.274 & 0.999 & 0.277 & 0.212 & -0.018 & 0.051 & -0.087 & 10.654 \\
   '' & 3.8211 & 6.2444 & 13878.8520 & -0.389 & 0.434 & 0.287 & 1.057 & 0.289 & 0.222 & -0.012 & 0.057 & -0.089 & 10.811 \\

\\
 0.160 & 3.9504 & 5.0327 & 1727.0790 & -0.185 & 0.153 & 0.066 & 0.149 & 0.071 & 0.070 & -0.088 & -0.001 & -0.103 & 6.418 \\
   ''  & 3.9698 & 5.2393 & 1945.7380 & -0.191 & 0.118 & 0.038 & 0.052 & 0.046 & 0.057 & -0.094 & -0.007 & -0.154 & 6.790 \\
   ''  & 3.9781 & 5.3928 & 2146.9080 & -0.206 & 0.109 & 0.030 & 0.017 & 0.038 & 0.053 & -0.096 & -0.009 & -0.181 & 7.118 \\
   ''  & 3.9824 & 5.5305 & 2354.0330 & -0.221 & 0.110 & 0.027 & 0.004 & 0.035 & 0.051 & -0.097 & -0.010 & -0.198 & 7.437 \\
   ''  & 3.9829 & 5.6229 & 2519.6030 & -0.234 & 0.116 & 0.029 & 0.005 & 0.037 & 0.051 & -0.098 & -0.011 & -0.201 & 7.666 \\
   ''  & 3.9796 & 5.7301 & 2756.9130 & -0.254 & 0.131 & 0.038 & 0.024 & 0.043 & 0.053 & -0.098 & -0.012 & -0.190 & 7.956 \\
   ''  & 3.9728 & 5.8230 & 3020.1000 & -0.276 & 0.149 & 0.051 & 0.064 & 0.055 & 0.058 & -0.097 & -0.013 & -0.173 & 8.240 \\
   ''  & 3.9631 & 5.9049 & 3311.5210 & -0.298 & 0.171 & 0.069 & 0.123 & 0.072 & 0.067 & -0.094 & -0.012 & -0.156 & 8.524 \\
   ''  & 3.9513 & 5.9786 & 3642.6880 & -0.322 & 0.194 & 0.090 & 0.198 & 0.093 & 0.078 & -0.088 & -0.009 & -0.139 & 8.809 \\
   ''  & 3.9384 & 6.0476 & 4027.5130 & -0.348 & 0.217 & 0.111 & 0.282 & 0.115 & 0.090 & -0.082 & -0.006 & -0.126 & 9.097 \\
   ''  & 3.9245 & 6.1127 & 4488.6580 & -0.374 & 0.242 & 0.134 & 0.374 & 0.137 & 0.105 & -0.075 & -0.002 & -0.116 & 9.389 \\
   ''  & 3.9137 & 6.1604 & 4945.4780 & -0.392 & 0.260 & 0.152 & 0.445 & 0.155 & 0.117 & -0.069 &  0.002 & -0.109 & 9.610 \\
   ''  & 3.9029 & 6.2081 & 5650.0730 & -0.408 & 0.278 & 0.169 & 0.516 & 0.171 & 0.129 & -0.063 &  0.007 & -0.104 & 9.831 \\
   ''  & 3.8887 & 6.2721 & 7814.6150 & -0.424 & 0.303 & 0.190 & 0.610 & 0.193 & 0.145 & -0.054 &  0.013 & -0.099 & 10.129 \\
   ''  & 3.8717 & 6.3256 & 11089.8240 & -0.433 & 0.333 & 0.215 & 0.723 & 0.217 & 0.165 & -0.044 &  0.022 & -0.094 & 10.429 \\
   ''  & 3.8577 & 6.3607 & 13851.5450 & -0.432 & 0.359 & 0.235 & 0.814 & 0.237 & 0.181 & -0.036 &  0.030 & -0.091 & 10.652 \\
   ''  & 3.8430 & 6.3929 & 16857.7690 & -0.423 & 0.387 & 0.256 & 0.912 & 0.258 & 0.198 & -0.027 &  0.039 & -0.090 & 10.879 \\
\\
\hline
\end{tabular}
\medskip
              
\end{minipage}
\end{table*}

\bsp

\label{lastpage}

\end{document}